\documentclass[11pt,twoside]{article} \usepackage[unicode]{hyperref}
 \newcommand\aut {Leonid A.~Levin} \newcommand\ttl {How do humans succeed
 in tasks like proving Fermat's Theorem or predicting the Higgs boson?}
  \usepackage [T1,T2A]
 {fontenc} \usepackage[utf8]{inputenc}\usepackage[russian,english]{babel}
 \usepackage[margin=1in,centering]{geometry} 
\begin{document} \frenchspacing

 \newcommand\hreff[1] {{\footnotesize\href{https://#1}{https://#1}}}
 \title {\ttl\thanks {This article is based on a talk at STOC-2021
 \url{http://acm-stoc.org/stoc2021/STOCprogram.html}\newline %breaks hunspell
 The talk video is available at \hreff{www.cs.bu.edu/fac/lnd/expo/stoc21/}
   \newline (and at \hreff{www.youtube.com/watch?v=8-x1uIGboNc} ).
   The whole 6/23 morning session \newline (my part: minutes 27-46)
 is at \hreff{www.youtube.com/watch?v=UgGGXXkYqsM}}}
 \date{}\author {\aut\thanks {Boston University, College of Arts
 and Sciences, Computer Science department, Boston, MA 02215.}}

 \maketitle\begin{abstract} I discuss issues of inverting
 feasibly computable functions, optimal\\ discovery algorithms,
 and the constant overheads in their performance. \end{abstract}

Our computers do a huge number of absolutely wonderful things.\\ Yet most of
these things seem rather mechanical. Lots of crucial problems that do\\ yield
to the intuition of our very slow brains are beyond our current computer arts.

Great many of these tasks can be stated in the form of\\ inverting easily
computable functions, or reduced to this form.\\ (That is, finding
inputs/actions that could produce a given result in a given realistic process.)

We have no idea about intrinsic difficulty of these tasks. And yet, traveling
salesmen do get to their destinations, mathematicians do find proofs of their
theorems, and physicists do find patterns in transformations of their bosons
and fermions !\\How is this done, and how could computers emulate their
success?

Of course, these are collective achievements of many~minds engrossed in a huge
number of papers. But today's computers can easily search through all math and
physics papers ever written. The limitation is not in physical capacity.

And brains of insects solve problems of such complexity and with such
efficiency, as we\\ cannot dream of. Yet, few of us would be flattered by
comparison to the brain of an insect.\\ What advantage do we humans have ?

\hspace{-1pt}One is the ability to solve {\bf new} problems on which evolution
did not train zillions of our ancestors. We must have some pretty universal
methods, not dependent on the specifics of focused problems.\\ Of course, it is
hard to tell how, say, mathematicians find their proofs. Yet, the diversity and
dynamism of math achievements suggest that some pretty universal mechanisms
must be at work.

Let me get now more technical, and focus on a specific problem:\\
Consider, for instance, algorithms that 3-color given graphs\footnote
 {This is a complete problem, i.e. all other
   inversion problems are reducible to it.}.\\
 Is it true that every such algorithm can be sped-up
 10 times on {\bf some} infinite set of graphs ?

 {\em {\bf\large Or,} there is a ``perfect'' algorithm, that
 cannot be outsped 10 times even on a {\bf subset} of graphs~?}

Note that there is a 3-coloring algorithm that cannot be outsped by more than\\
a constant factor on {\bf any} subset. The question is, must this constant get
really big ?

\subsection*{***}\vspace{-10pt}

But before further discussion, let me go into some history.

In the 50s, in the Russian math community there was much interest in the works
of Claude Shannon. But many of Shannon's constructions required exhaustive
search of all configurations. There was an intense interest in whether these
exponential procedures could be eliminated (see \cite{trakh}).

And Sergey Yablonsky wrote a paper that he interpreted as showing that no\\
subexponential method could work on a problem that is, in today's terms, co-NP.
\\It is a problem of finding a boolean function of maximal circuit complexity.

Kolmogorov saw this claim as baseless since the proof considered only
a specific\\ type of algorithms. He was unhappy with such a misleading idea
being promoted.\\ Kolmogorov advocated the need for efforts to find valid proofs
that some\\ commonly believed complexities of popular problems are, in fact,
unavoidable.

This required a convincing definition of the running time. But Turing Machines
were seen\\ as too restricted to use for meaningful speed lower bounds.
Kolmogorov formulated (see \cite{ku})\\ a graph-based model of algorithms that
had time complexities as we understand them today.

He also ran a seminar where he challenged mathematicians with quadratic
complexity\\ of multiplication. And an unexpected answer was soon found by
Anatoly Karatsuba, and\\ improved by Andrei Toom: multiplication complexity
turned out to be nearly linear.\\
 (It is now really fast with subsequent improvements by Cook and others !)

This was an impressive indication that common sense is an unreliable guide\\
 for hardness of computational problems, and must be verified by valid proofs.

\vspace{-9pt}\subsection*{***}\vspace{-10pt} 

I, at that time, was extremely excited by some other work of Kolmogorov.\\
 He (and independently Ray Solomonoff) used the Turing's Universal Algorithm
for an\\ optimal definition of informational complexity, randomness, and some
other related concepts.

I noted that similar constructions yield an optimal up to a constant
factor\\ algorithm for a problem now called Tiling, and therefore for any search
problem,\\ as they all have a straightforward reduction to Tiling.

To my shagreen, Kolmogorov was not impressed with the concept of optimality,
saw it as too abstract for the issue at hand. (Indeed, finding specific bounds
did not look as hopeless then as it now does.) But he was much more interested
in my remark that Tiling allows reduction to it of all\\ other search problems.
He thought I should publish {\bf that} rather than the optimal search.

I thought it would only be worth publishing if I can reduce it to some popular
problems.\\ My obstacle was that combinatorics was not popular in Russia, and
my choice\\ of problems that might impress the math community was rather
limited.\\ I saw no hope for something like factoring, but spent years in naive
attempts\\ on things like graph isomorphism, finding small circuits for boolean
tables, etc.

Meanwhile an interesting angle was added to the issues. In 1969 Michael
Dekhtiar, a student\\ of Boris Trakhtenbrot, published a proof \cite{d} that
under some oracles inverting simple functions\\ has exponential complexity.
In the US, Baker, Gill, and Solovay did this independently \cite{bgs}.

Later I ran into problems with communist authorities. And friends advised me to
quickly publish all I have while the access to publishing is not yet closed to
me. So I submitted several papers in that 1972, including the one about search
\cite{ll} (where Kolmogorov agreed to let me include the optimal search).
I guess I must thank the communists for this publication.

But the greatest developments by far were going on in the United States.\\
Cook \cite{cook}, Karp\cite{karp}, and Garey and Johnson \cite{gj}
 made a really revolutionary discovery.\\
 They found that 3-SAT reduces to great many important combinatorics problems.

Combinatorics received much attention in the West
 and these results became a coup !

\subsection*{***}

Kolmogorov asked several questions at that time, still open and
interesting.\\ One was: Are there polynomial time algorithms that have no
\mbox{\bf linear} size circuits ?\\ We knew that some slow polynomial time
algorithms cannot be replaced by faster {\bf algorithms}.\\
 But can linear-sized circuits families replace {\bf all} of them ?

His other interesting comment was a bit more involved. We proved at that time
that mutual information between strings is roughly symmetric. The proof
involved exponential search for\\ short programs transforming a strings $x$ into
$y$. Kolmogorov wondered if such search for short\\ fast (meant in robust terms,
tolerating $+O(1)$ slacks in length and in $\log$ time) programs would\\
not be a better candidate than my Tiling to see if search problems are
exponentially hard.

He said that, often, a good candidate to consider is one that is neither too
general, nor too narrow. Tiling, being universal, may be too general, lacking
focus. Some other problems (say, factoring) -- too narrow. And search for fast
short programs looked like a good middle bet to him.\\ It still does to me ! :-)

Such search is involved in another type of problems that challenge our
creativity: extrapolating the observed data to their whole natural domains.
 It is called by many names, ``Inductive Inference'', ``passive learning'', and
others.
 Occam Razor is a famous principle of extrapolation. A version attributed to
Einstein suggests: hypothesis need be chosen as simple as possible, but no
simpler~:-).

\subsection*{***}

Ray Solomonoff gave it a more formal expression: ~~ The likelihoods of various
extrapolations, consistent with known data, decrease exponentially with the
length of their shortest descriptions. Those short programs run about as fast
as the process that had generated the data.

There have been several technical issues that required further attention. I
will stay on a simple side, not going into those details. Most of them have
been clarified by now, {\bf if} we ignore the time needed to find such short
fast programs. This may be hard. Yet, this is still an inversion task,
bringing us back to the issues of optimal search. I have a little discussion of
such issues in \cite{ll13}.

\subsection*{***}

Now, back to my focus. The concept of optimal algorithm for search problems
ignores\\ constant factors {\bf completely}. So, it is tempting to assume that
they must be enormous.

However, this does not seem so to me. Our brains have evolved on jumping in
trees, not\\ on writing math articles. And yet, we prove Fermat's Theorems,
design nukes, and even write\\ STOC papers. We must have some quite efficient
and quite universal guessing algorithms built-in.

So, I repeat a formal question about these constants:

 \begin{center} {\bf Can every algorithm for complete search problems\\
    be outsped 10 times on an infinite subset ?\\ {\Large OR,}
    there is a ``perfect''  one that cannot be, even on a subset ?}\end{center}

\noindent Of course, careless definitions of time can allow fake speed-ups. For
instance if we ignore the alphabet size and reduce the number of steps just by
making each step larger due to the larger alphabet. Or if we exclude the
required end testing of the input/output relation, and choose a relation that
itself allows a non-constant speed-up. But it is easy to carefully define time
to preclude such cheating.

\subsection*{***}

Let me now go into some little technicalities to see what issues are involved
in understanding\\ these constant factors. We look at the optimal search for an
inverse $w$ of a fast algorithm $f$,\\
 given the output $x$ that $f$ must produce from $w$.

 We refine Kolmogorov Complexity with {\bf time}, making it computable.\\
 The time-refined complexity {\bf Kt} of $w$ given $x$ considers all prefixless
programs $p$ by which\\ the universal algorithm $U$ generates $w$ from $x$ in
time $T$. That time also includes\\ running $f(w)$ to confirm it
is $x$. {\bf Kt$(w|x)$} is the minimum of the length of $p$, plus $\log T$.

The Optimal Inverter searches for solutions $w$ in increasing order of
this complexity {\bf Kt} of\\ $w$ given $x$, {\bf not} of length of $w$.
For instance, shorter proofs may be much harder to find, having\\
 higher complexities.
 The Inverter generates and checks in time $2^k$ all $w$ up to complexity $k$.

Btw, the optimal search makes the concept of complexity applicable to
individual instances\\ of search tasks, not just to families of instances which
we now call ``problems'' and\\ complexities of which we study. So we can ask how
hard is, say, to find a short proof\\ for Fermat's theorem, not for theorems in
general. Would not this notion fit tighter ?

The big {\bf catch} here is that {\bf each} wasteful bit $U$ requires of
$p$ {\bf doubles} the time. We would need a {\bf very} ``pure'' $U$, frugal
with wasting bits. Do our brains have such a one built-in ? It seems so\\
to me. We do seem to have little disagreement on what is ``neat'' and what is
cumbersome.\\ There are differences in our tastes, but they are not so huge
that we could not understand\\ each other's aesthetics.
 But this is just a feeling. The formal question remains:

 {\bf Is there an algorithm for a complete search problem that\par
 cannot be outsped ten times, even on an infinite subset ?}\\
 (Of course, this 10 is a bit arbitrary, can be
 replaced with your favorite reasonable constant.)

  \end{document}